\documentclass[a4paper,accepted=2024-07-10]{quantumarticle}
\pdfoutput=1
\usepackage[numbers,sort&compress]{natbib}

\usepackage[utf8]{inputenc}
\usepackage[english]{babel}
\usepackage[T1]{fontenc}
\usepackage{hyperref}

\usepackage{amsmath,amssymb,amsthm}
\usepackage{graphicx}%
\graphicspath{{figs/}}
\usepackage{dcolumn}%
\usepackage{enumitem}
\usepackage[normalem]{ulem}
\usepackage{float}
\usepackage{bm,dsfont}%
\usepackage{multirow}
\usepackage{svg}
\usepackage{physics}
\usepackage{color}
\usepackage{qcircuit}
\usepackage[caption=false]{subfig}

\hypersetup{
	colorlinks = true,
	linkcolor = [rgb]{0.70,0.13,0.13},
	citecolor = [rgb]{0.13,0.55,0.13},
	urlcolor = [rgb]{0.25, 0.41, 0.88}}
 \usepackage[capitalize]{cleveref}
\usepackage{algorithm}
\usepackage{algpseudocode}

\newtheorem{theorem}{Theorem}[section]

\newtheorem{definition}{Definition}[section]
\newcommand{\ii}{\mathrm{i}}

\newcommand{\scB}{\mathcal{B}}

\newcommand{\scD}{\mathcal{D}}
\newcommand{\scE}{\mathcal{E}}

\newcommand{\scO}{\mathcal{O}}
\newcommand{\scT}{\mathcal{T}}
\newcommand{\scS}{\mathcal{S}}

\begin{document}

\title{Zero and Finite Temperature Quantum Simulations Powered by Quantum Magic}
\author{Andi Gu}
\thanks{Equal contributions (Alphabetical order)}
\affiliation{Department of Physics, Harvard University, 17 Oxford Street, Cambridge, MA 02138, USA}

\author{Hong-Ye Hu}
\thanks{Equal contributions (Alphabetical order)}
\affiliation{Department of Physics, Harvard University, 17 Oxford Street, Cambridge, MA 02138, USA}

\author{Di Luo}
\thanks{Equal contributions (Alphabetical order)}
\affiliation{Department of Physics, Harvard University, 17 Oxford Street, Cambridge, MA 02138, USA}
\affiliation{Center for Theoretical Physics, Massachusetts Institute of Technology, Cambridge, MA 02139, USA}
\affiliation{The NSF AI Institute for Artificial Intelligence and Fundamental Interactions}
\author{Taylor L. Patti}
\affiliation{NVIDIA, Santa Clara, CA 95051, USA}
\author{Nicholas C. Rubin}
\affiliation{Google Quantum AI, Venice, CA 90291, United States}
\author{Susanne F. Yelin}
\email{syelin@g.harvard.edu}
\affiliation{Department of Physics, Harvard University, 17 Oxford Street, Cambridge, MA 02138, USA}
\affiliation{The NSF AI Institute for Artificial Intelligence and Fundamental Interactions}

\begin{abstract}
We introduce a quantum information theory-inspired method to improve the characterization of many-body Hamiltonians on near-term quantum devices. We design a new class of similarity transformations that, when applied as a preprocessing step, can substantially simplify a Hamiltonian for subsequent analysis on quantum hardware. By design, these transformations can be identified and applied efficiently using purely classical resources. In practice, these transformations allow us to shorten requisite physical circuit-depths, overcoming constraints imposed by imperfect near-term hardware. Importantly, the quality of our transformations is \emph{tunable}: we define a `ladder' of transformations that yields increasingly simple Hamiltonians at the cost of more classical computation. Using quantum chemistry as a benchmark application, we demonstrate that our protocol leads to significant performance improvements for zero and finite temperature free energy calculations on both digital and analog quantum hardware. Specifically, our energy estimates not only outperform traditional Hartree-Fock solutions, but this performance gap also consistently widens as we tune up the quality of our transformations. In short, our quantum information-based approach opens promising new pathways to realizing useful and feasible quantum chemistry algorithms on near-term hardware.
\end{abstract}

\maketitle

A central task in quantum chemistry is determining ground state energies and finite temperature free energies for electronic Hamiltonians. While numerous algorithms aim to leverage quantum hardware for solutions \cite{VQE,Yuan2019theoryofvariational,2019arXiv190911097Y,QPE}, the constraints of near-term hardware, particularly limited circuit depth, pose challenges. One approach to address this difficulty is classically assisted quantum algorithms, wherein computation is offloaded to a classical computer where possible, and the quantum hardware only supplies the quantum resources which are absolutely necessary to identify an accurate solution. A natural way in which a classical computer might assist a quantum algorithm is to preprocess the electronic Hamiltonian into a more favorable form before it is analyzed on quantum hardware. For instance, one can use a classical algorithm to partition the terms of a Hamiltonian $H$ into commuting groups which can be measured simultaneously, so that calculating the expectation $\expval{H}$ can be done with fewer samples \cite{verteletskyi2020}. 

In this work, we consider another preprocessing method: \emph{similarity transformations}, wherein our aim is to identify a unitary change of basis $U$ such that the transformed Hamiltonian $H_{\text{eff}} \equiv U^\dagger H U$ is more amenable to analysis on near-term quantum hardware. Indeed, similarity transforms have long been used to simplify Hamiltonians~\cite{PhysRevB.93.104205,PhysRevB.103.L100207,PhysRevB.94.014205,PhysRevLett.116.247204,PhysRevB.94.045111,roussy_approximate_1973,cederbaum_block_1989,bravyi_schriefferwolff_2011,white_numerical_2002}. In this work, we take a novel quantum information theoretic approach to similarity transforms: we center our techniques around two fundamental measures of `quantumness' -- namely, entanglement and nonstabilizerness (popularly termed `quantum magic'). These notions of quantumness formalize the intuition that the degree to which a state is truly quantum can be measured by how difficult it is to simulate classically. For instance, low entanglement states can be simulated by matrix product states \cite{cirac2021}. However, entanglement alone is insufficient to completely characterize quantumness -- classically simulable (i.e., using classical resources that scales polynomially in system size) circuits \cite{PhysRevA.70.052328}, known as Clifford circuits can generate an unbounded amount of entanglement. Indeed, it has been shown that random Clifford circuits will almost always produce volume-law entangled states~\cite{webb2016clifford}. Yet, these circuits are unable to express a different quantum resource: quantum magic. More precisely, Clifford circuits are generated by a non-universal gateset (composed of the Hadamard, phase, and CNOT gates); this has motivated the formal definition of magic as a measure of the amount of non-Clifford resources required to prepare a particular state~\cite{howard2017application,seddon2019quantifying}. Like entanglement, quantum magic has also been shown to be directly connected with the hardness of classical simulation~\cite{Bravyi_2019}.

\begin{figure}[ht]
    \centering
    \includegraphics[width=\columnwidth]{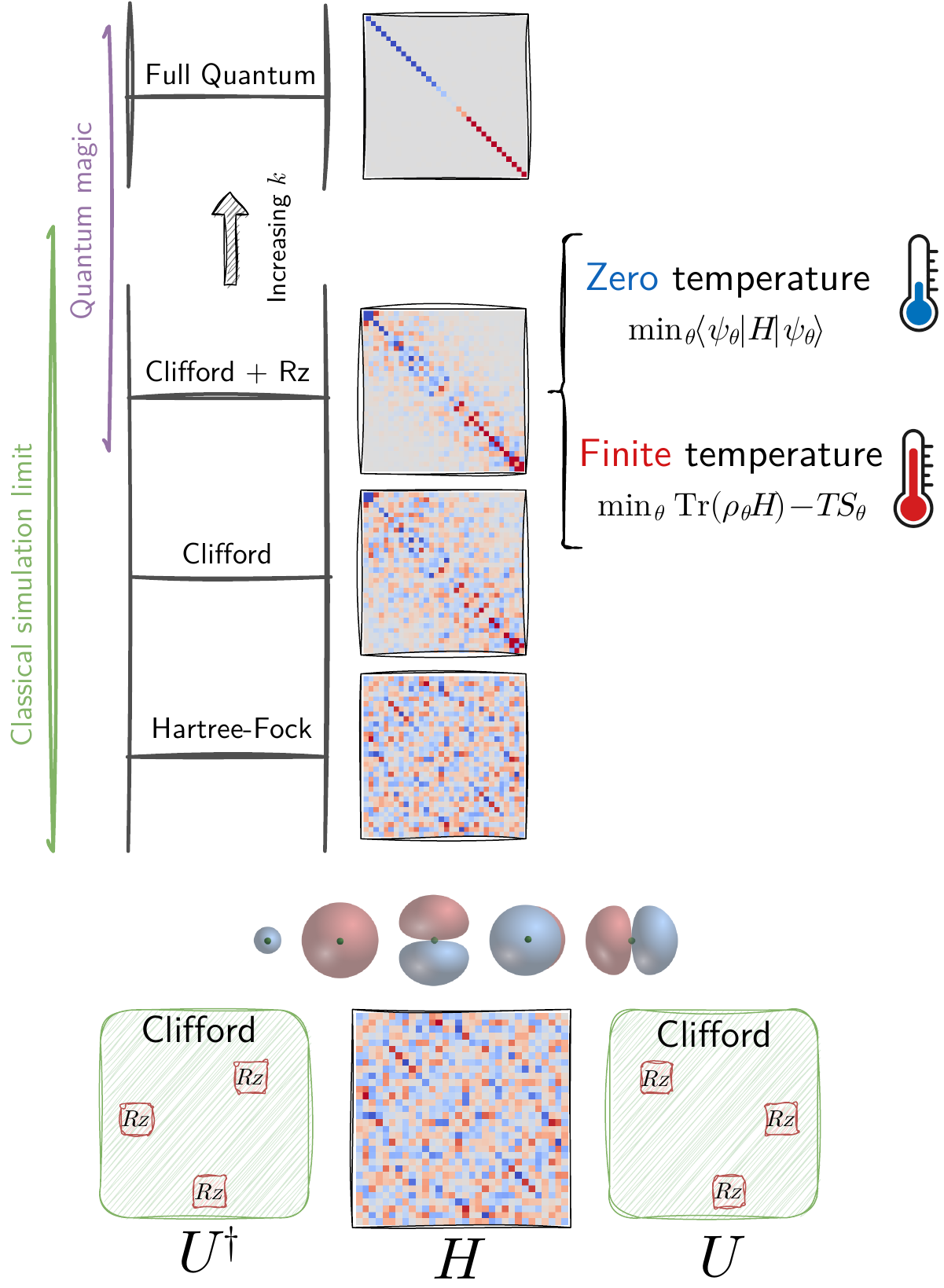}
    \caption{A general approach for similarity transformations using the quantum magic ladder. Starting with a second quantized Hamiltonian $H$, we use classical simulable quantum circuits (e.g., Clifford + $k$Rz circuits) to reduce the quantumness of the Hamiltonian by the basis transformation $H_{\text{eff}}=U^{\dagger}H U$. We climb up the quantum magic ladder by increasing $k$, which increases the degree to which $U$ can simplify the Hamiltonian (by reducing the quantum magic of the $H_{\text{eff}}$ eigenbasis). Note that for finite temperature calculations, when working in the grand canonical ensemble, we replace $H$ with $H-\mu N$.}
    \label{fig:theme}
\end{figure}

Our main insight in this work is that similarity transforms can be used to eliminate spurious quantumness (i.e., classically simulable entanglement and magic) in the Hamiltonian's eigenbasis before it is analyzed on quantum hardware. Then, the real hardware only needs to supply the fundamentally necessary quantumness for the problem, hence helping shorten requisite circuit depths. We introduce a class of similarity transformations, represented as Clifford circuits augmented with a few (non-Clifford) single-qubit $Z$-rotation gates. For brevity, we call these circuits Clifford + $k$Rz circuits. Conveniently, this design allows good similarity transforms to be both identified and applied \emph{efficiently using purely classical resources}. The Clifford component of the circuit can be interpreted as eliminating spurious entanglement from $H_{\text{eff}}$, while the non-Clifford gates eliminate spurious magic. Importantly, by varying the number of non-Clifford gates, we show that the power of these transformations is \emph{tunable}, so that we can identify increasingly simple eigenbases (i.e., exhibiting less entanglement and magic) at the cost of increased classical computational resources. Since this tunability comes in discrete steps, we term this the ``quantum magic ladder'' (see \cref{fig:theme}). By applying our approach to zero and finite temperature free energy calculations, we demonstrate that this reduction in quantumness can have significant practical benefits. Specifically, we show that our similarity transforms find energies which are significantly more accurate than other popular classical solutions, and moreover, the performance gap only improves as we climb the quantum magic ladder. Furthermore, we present a classical algorithm which outputs a new (simplified) effective Hamiltonian $H_{\text{eff}}$, related to the original Hamiltonian via a similarity transform. We show that the resultant simplified Hamiltonians $H_{\text{eff}}$ produced by our method enables significantly reduced circuit depths on quantum hardware. Under this quantum information theoretic lens, these results demonstrate a promising pathway for designing practical quantum chemistry algorithms in the NISQ era.

\textit{Quantum Chemistry.} Quantum chemistry is a natural domain for assessing the efficacy of our protocol. The Hamiltonians produced by quantum chemistry systems have immediate physical interpretations, allowing us to apply physical intuition to better understand how our protocol modulates the `quantumness' of a problem. Furthermore, there is a wealth of established quantum chemistry methods against which we can benchmark our protocol, such as the Hartree-Fock (HF) approximation. In short, quantum chemistry allows us to demonstrate the advantages of our approach in a well-understood physical context.

In quantum chemistry, one key goal of the electronic structure problem is to solve the energy of many-body electronic Hamiltonians at zero and finite temperature. These electronic Hamiltonians are often expressed in a molecular orbital basis:
\begin{equation}
    H = \sum_{p,q} h_{pq}\hat{c}^{\dagger}_{p}\hat{c}_{q}+\dfrac{1}{2}\sum_{p,q,r,s}V_{pqrs}\hat{c}^{\dagger}_{p}\hat{c}^{\dagger}_{q}\hat{c}_{s}\hat{c}_{r},\label{eq:second_q}
\end{equation}
where $p,q,r,s$ label the spin-orbitals and $\hat{c}_{\alpha}$, $\hat{c}^{\dagger}_{\alpha}$ are the annihilation and creation fermionic operators. Anticipating analysis on qubit-based quantum computers, we apply a fermion-qubit mapping (e.g., parity mapping \cite{10.1063/1.4768229}, Jordan-Wigner \cite{Jordan:1928wi}, Bravyi-Kitaev \cite{https://doi.org/10.1002/qua.24969}) to  map the fermionic Hamiltonian to a qubitized Hamiltonian $H = \sum_{i}\alpha_i P_i$, where $P_i$ is an $n$-qubit Pauli operator. For the molecules considered in this work (LiH and H$_2$O), the qubitized Hamiltonians contain $O(10^2)$ Pauli strings. The explicit construction of these Hamiltonians was performed using the \texttt{OpenFermion} library~\cite{mcclean2019openfermion}. In this work, we focus on two paradigmatic problems involving these electronic Hamiltonians, which are to calculate the ground state energy of $H$, and the free energy of the Gibbs state at finite temperature \cite{chowdhury2016quantum}. In this work, we work in the grand canonical ensemble, so that the Gibbs state is $\rho \propto \exp(-\beta (H - \mu N))$, with inverse temperature $\beta$, chemical potential $\mu$, and particle number $N$.

\textit{The Quantum Magic Ladder.} Similarity transformations have historically occupied a key role in the development of quantum mechanics. An early example is the Schrieffer-Wolff method, which generates low-energy effective Hamiltonians by constructing similarity transformations that separate high-energy and low-energy subspaces \cite{roussy_approximate_1973,cederbaum_block_1989,bravyi_schriefferwolff_2011,white_numerical_2002}. In keeping with this tradition, we argue that similarity transforms can also play a central role in bringing quantum chemistry algorithms closer to practical feasibility in the NISQ era. One way of understanding the role of a similarity transformation $U$, particularly in the context of near-term hardware, is as a virtual circuit that is appended to the end of a ground state preparation circuit $V$. That is, if we define a transformed Hamiltonian $H_{\text{eff}} = U^\dagger H U$, we observe that $\expval{V^\dagger H_{\text{eff}} V} = \expval{(UV)^\dagger H (U V)}$, so that, in effect, the applied circuit is $UV$. This is especially important for NISQ devices, which face limited circuit depths (hence limited expressive power of $V$). An appropriate similarity transformation $U$ can act like an appended circuit that significantly eases the burden of the real quantum circuit $V$. 

In what follows, we describe a unified framework to programatically construct a new class of similarity transformations for quantum chemistry algorithms, a system we term ``the quantum magic ladder.'' These transformations are comprised of Clifford circuits augmented by a few Rz gates. There are several reasons for the attractiveness of the Clifford + $k$Rz approach. Firstly, Clifford circuits are classically simulable with complexity $O(n^3)$, where $n$ is the system size \cite{PhysRevA.70.052328}. As we discuss below, these Clifford circuits can be dramatically enhanced by doping them with just a few ($k$) non-Clifford gates. While this enhancement comes at the cost of an increase in simulation complexity that is exponential in $k$, Clifford + $k$Rz transforms with $k \leq O(\log n)$ can still be simulated with polynomial classical resources \cite{PhysRevA.70.052328,PhysRevLett.116.250501,PhysRevLett.128.220503}. Secondly our circuits strictly generalize more conventional circuits such as Clifford circuits with a few T gates. Indeed, we find that the additional continuous degrees of freedom introduced by Clifford + $k$Rz circuits (i.e., the rotation angle of each Rz gate) is both easy to optimize and yields significant performance improvements. Furthermore, these improvements are free in the sense that the additional simulation cost for $k$Rz gates is exactly the same as that of $k$ T gates. The simulation of these Clifford + $k$Rz circuits was performed using the \texttt{PyClifford} package \cite{PyClifford} (for details, see \cref{app:general-simulation}). Finally, as illustrated in \cref{fig:theme}, `climbing' the quantum magic ladder provides a \emph{systematic} way to increase the expressive power of our similarity transformation: one simply needs to increase $k$.

The enhanced expressivity of ansatze higher on this ladder can be understood through the lens of quantum magic. Much like entanglement defines one notion of quantumness, quantum magic, or nonstabilizerness, is another metric which measures the degree to which a state can be considered truly quantum~\cite{howard2017application,seddon2019quantifying}.  Our main insight in this work is that the optimal Clifford + $k$Rz transform can model the quantumness of $H$ along \textit{both} axes of quantumness, capturing its entanglement via the Clifford component of the circuit, and its quantum magic via the non-Clifford components. As we climb the magic ladder, we increase the degree to which our transform can model the quantum magic of the energy eigenbasis (hence reducing the magic of the transformed Hamiltonian's eigenbasis). In the limit as $k \rightarrow \infty$, our ansatz can model any unitary \cite{Nebe2006}, which ultimately leads to a full reduction in the Hamiltonian's quantumness (i.e., complete diagonalization). 

Our Clifford + $k$Rz circuits are parameterized as $U(\Theta)$, where $\Theta$ is a combination of discrete and continuous parameters, corresponding to the Clifford and Rz components respectively. To identify good transformations $U(\Theta)$, we make use of the variational principle for both ground state energies and finite temperature free energies. Specifically, consider the zero temperature case: $U(\Theta)$ defines a parameterized effective Hamiltonian $H_{\text{eff}}(\Theta)$, and our aim is then to minimize $\expval{H_{\text{eff}}(\Theta)}{0}$ (which is equivalent to minimizing the energy, with respect to $H$, of $U(\Theta) \ket{0}$). Optimizing $\Theta$ is a hybrid discrete-continuous problem with a large search space, so we develop a three-step approach. First, we observe that qubitized electronic Hamiltonians strongly resemble disordered Hamiltonians, whose energy coefficients span a wide range of scales. We leverage this formal similarity by using the spectrum bifurcation renormalization group \cite{PhysRevB.93.104205,PhysRevB.103.L100207,PhysRevB.94.014205} (which produces Clifford similarity transformations effective for modelling strongly-disordered quantum materials) to find an initialization point $\Theta_0$ for our Clifford + $k$Rz transformation. We then perform a combination of greedy search and gradient descent to further optimize $\Theta$ (see \cref{app:optimization} for details). 

\begin{figure}[ht]
    \centering
    \includegraphics[width=0.9\columnwidth]{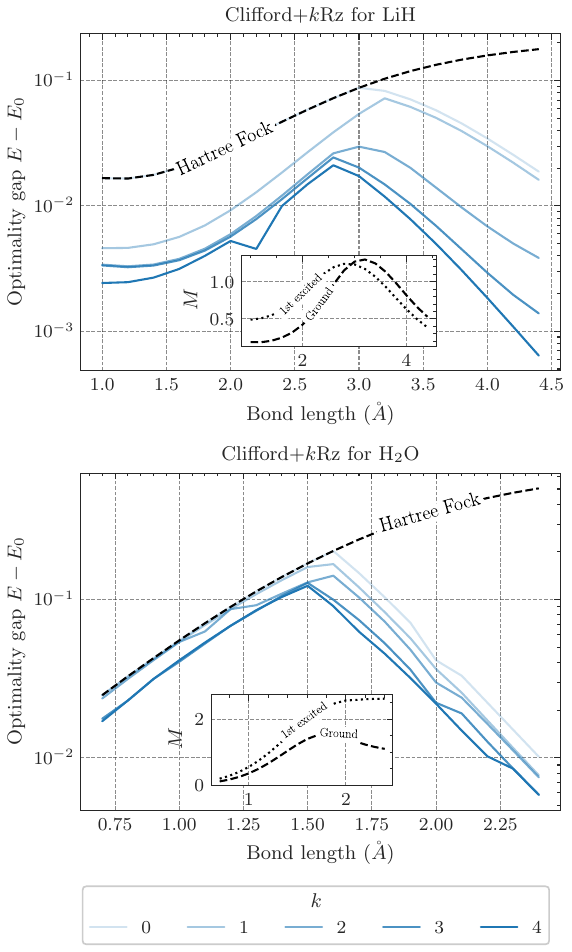}
    \caption{$E-E_0$ is the gap between $E$, the energy calculated by the optimized Clifford + $k$Rz circuit, and $E_0$, the true ground state energy. We illustrate this gap for LiH and H$_{2}$O as a function of their bond lengths. Note that the HF solution is consistently higher in energy, as it is strictly less expressive than the Clifford ansatz, even when $k=0$, underscoring the enhanced representational power of the Clifford circuits in capturing the ground state properties. Furthermore, the accuracy (as measured by $E-E_0$) improves as we climb the quantum magic ladder. The inset shows the stabilizer R\'enyi entropy $M$, which quantifies the magic of the ground and first excited states at different bond lengths.}
    \label{fig:lih}
\end{figure}

\textit{Results.} We apply our Clifford+\textit{k}Rz transformation for both ground state and finite temperature quantum chemistry calculations~\cite{zhang2010impacts,RevModPhys.88.045005,DONOGHUE1985233,sun2021quantum,powers2023exploring} on LiH ($n=8$) and H$_{2}$O ($n=12$). While these system sizes are modest compared to the capabilities of state-of-the-art classical numerical methods, they serve as valuable proof-of-concept demonstrations for our quantum-classical hybrid approach, and are representative of the system scales that are expected to be tractable on near-term quantum devices. We start with the zero temperature case, where we aim to identify a ground state energy $E=\expval{H_{\text{eff}}(\Theta)}{0}$ that closely approximates the true ground state energy $E_0$. In \cref{fig:lih}, the energy gap $E-E_0$ is plotted as a function of different bond lengths between atoms, where $E$ is the energy predicted by an optimized from Clifford + $k$Rz transformation. First, compare the accuracy to the accuracy of the Hartree-Fock (HF) solution: since the HF solution is a strict subset of the Clifford ansatz~\cite{ravi2022cafqa}, the energy gap of a pure Clifford transformation ($k=0$) is upper bounded by that of the HF solution. However, a pure Clifford circuit only surpasses the HF solution when the bond length exceeds the equilibrium bond length. In contrast, as we climb the quantum magic ladder, the energy gap is significantly reduced for \emph{all} the bond lengths, including bond lengths shorter than the equilibrium bond length. Moreover, as the number of non-Clifford gates increases, the energy gap consistently decreases. This is clear evidence that tuning up the power of the Clifford + $k$Rz ansatz can significantly improve accuracy for ground state calculations. It is important to note that the number of non-Clifford gates $k$ should be considered in relation to the total number of qubits $n$. If we allow $k=\exp(\Theta(n))$, the doped circuit can approximate an arbitrary unitary gate. In contrast, in our study, we focus on the regime where $k \ll n$ to maintain the classical simulability of the circuits while still enhancing their expressibility. We observe that the convergence of our method improves as $k$ increases, with significant improvements seen for $k=1$ and $k=2$. However, the improvement becomes marginal for higher values of $k$. There is a trade-off between the expressibility gained by increasing $k$ and the computational cost associated with it; it appears that there are sharply diminishing returns beyond $k=2$. Future work could investigate this trade-off in more detail and develop strategies for optimally selecting $k$ based on the problem at hand. We also note the presence of kinks in some curves of \cref{fig:lih}, which are artifacts likely due to numerical precision limitations and minor optimization errors, rather than physical phenomena.

Interestingly, the energy gap peaks around a particular bond length, indicating a region of maximal quantum magic. To show this, we employ a popular metric specifically designed to quantify the magic of a state $\ket{\psi}$. This metric $M(\ket{\psi})$ is known as the stabilizer entropy~\cite{leone2022}, and is defined as
\begin{equation}
    M(\psi) = -\log_2\qty(\sum_{P \in \mathbb{P}_n} \frac{\tr^4(P \psi)}{2^n}),
\end{equation}
where $\mathbb{P}_n$ is the $n$-qubit Pauli group.
Notably, it obeys $M(\ket{\psi})=0$ if and only if $\ket{\psi}$ is a stabilizer state, and it is strictly positive otherwise. In general, a higher value of $M(\ket{\psi})$ indicates a greater degree of nonstabilizerness~\cite{leone2022}. We see in the inset of \cref{fig:lih} that quantum magic (as measured by $M$) and the energy gap peak near the same bond length, indicating that regions where the ground state has high quantum magic cannot accurately be reduced by basis transformations with insufficient non-Clifford gates. Interestingly, we observe that the magic of the ground state is typically very small at extreme bond lengths, and peaks at intermediate bond lengths. This peak in magic and energy gap likely corresponds to the Coulson-Fisher point~\cite{coulson1949}, where the electronic structure transitions from single-reference to multi-reference character. This transition point is known to be challenging for computational methods~\cite{limacher2013}, aligning with our observation of increased quantum resources being necessary in this region. Our quantum information theoretic approach thus provides a new perspective on this well-established concept in quantum chemistry. This suggests that in practice, hyperparameters like $k$ ought to be set accordingly --- for extreme bond lengths, a small $k$ should suffice, while intermediate bond lengths will require a larger $k$ to accurately capture the increased magic. Near these intermediate bond length regions, accurate results demand either more powerful classical similarity transformations (i.e., larger $k$) or further analysis on real quantum devices.

\begin{figure}
    \centering
    \includegraphics[width=0.9\columnwidth]{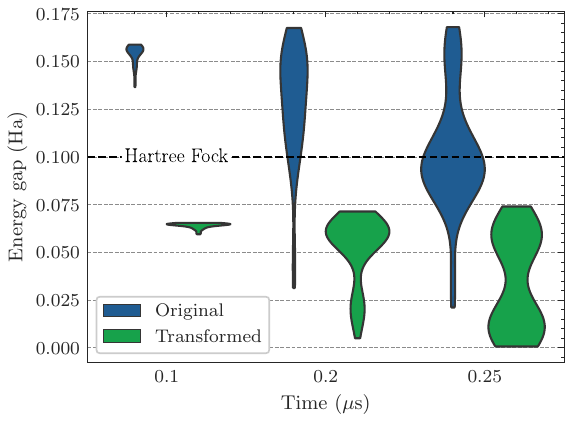}
    \caption{Pulse time reduction with the transformed Hamiltonian, depicted as a violin plot. This displays the probability density of a solution having a particular energy gap, with the width representing the solution density at each energy level. We model an analog quantum machine with a $^{87}$Rb atomic chain for LiH ground state preparation using fixed lattice spacing and total evolution time. The Rabi frequency and the local detuning of each atom are optimized. For each evolution time, twenty randomly initialized variational ground states are optimized.}
    \label{fig:pulse}
\end{figure}

\begin{figure}
    \centering
    \includegraphics[width=0.9\columnwidth]{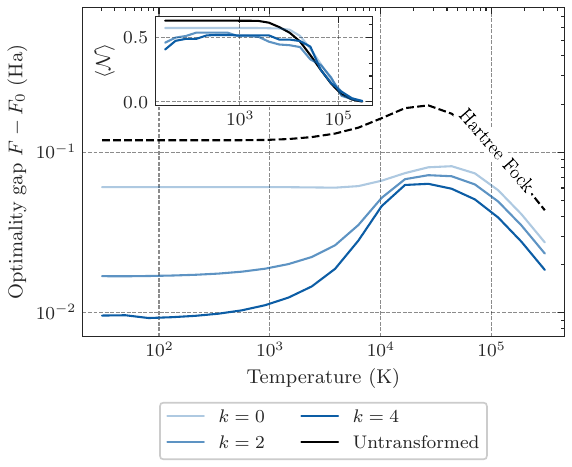}
    \caption{Free energy gap for LiH at a bond length 3.4\AA. The inset shows the entanglement negativity $\mathcal{N}$ averaged over each site of the Gibbs state for the transformed Hamiltonian.}
    \label{fig:fin-temp}
\end{figure}

We demonstrate this by applying our method on analog quantum computers. Specifically, we classically simulate the dynamics of a one-dimensional Rydberg array, which serves as a programmable quantum simulator (PQS) for preparing a ground state of LiH at a bond length 3.4\r{A}. Note that this bond length exhibits high magic (\cref{fig:lih}), indicating that additional quantumness, supplied by the PQS, is necessary to prepare the ground state. To prepare this state, we fix the total PQS evolution time $T$ and evolve the initial state $\ket{0}$ under a time-dependent Rydberg Hamiltonian \cite{Adams_2019} with variationally optimized site-dependent Rabi frequencies $\Omega_{i}(t)$ and local detunings $\Delta_i(t)$. For each fixed total evolution time $T$, we variationally optimize 20 randomly initialized ground states for both the bare Hamiltonian $H$, and for the Clifford-transformed Hamiltonian $H_{\text{eff}}$, using automatic differentiation with Jax \cite{jax2018github} and Tensorcircuit \cite{Zhang2023tensorcircuit} to directly optimize the pulse shapes $\Omega_i(t), \Delta_i(t)$. The results are shown in \cref{fig:pulse}. Consider just the case of $T=0.1\mu$s, for which dramatic differences are already clear. The optimized energy with the original Hamiltonian is concentrated at levels significantly above the HF energy, because the entanglement generated by the PQS within this small timeframe is limited. However, the Clifford + $k$Rz similarity transform already significantly reduces the entanglement of the ground state before evolution on the PQS; hence, even under the limited evolution time $T=0.1\mu$s, the PQS can quickly identify a solution that vastly surpasses the accuracy of the Hartree Fock solution. This demonstrates that Clifford + $k$Rz transformations effectively reduce requisite entanglement and pulse times, a significant advantage for near-term quantum devices.
 
For finite temperature calculations, we apply our approach to the LiH molecule for grand free energy estimation. Our ansatz for the density matrix is
\begin{equation}
    \rho(\Theta) = \sum_{x\in\{0,1\}^{n}}p(x)U(\Theta) \ketbra{x} U^\dagger(\Theta),
\end{equation}
where $U(\Theta)$ is a Clifford + $k$Rz circuit. Again applying the variational principle, we solve for $\Theta$ by minimizing the free energy $F=\langle H\rangle-\mu\langle N\rangle -T S$ of $\rho(\Theta)$ (see \cref{app:optimization} for details). The results in \cref{fig:fin-temp} show that the Clifford + $k$Rz ansatz consistently reduces the free energy gap $F-F_0$ (where $F$ is the free energy of the optimized density matrix $\rho(\Theta)$, and $F_0$ is the free energy of the true Gibbs state) as we climb the quantum magic ladder. Similar to the zero temperature case, all Clifford transformations outperform the finite temperature HF method, including the pure Clifford transformation ($k=0$), which reduces the free energy gap by $\sim 50\%$. 

\begin{figure}
    \centering
    \includegraphics[width=\columnwidth]{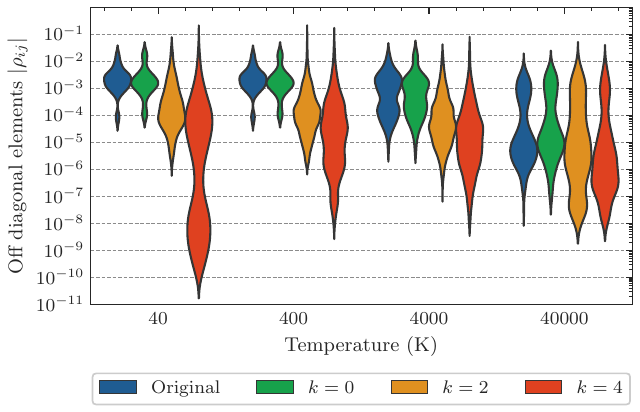}
    \caption{Density distribution of the off-diagonal term for the Gibbs state $\rho$ of LiH across different temperatures for bond length 3.4\AA. The width of features is proportional to solution density.}
    \label{fig:off-diag}
\end{figure}

To quantify the extent to which our transforms reduce the quantumness of the transformed Hamiltonian, we examine the entanglement features and off-diagonal terms of the exact Gibbs state for the transformed Hamiltonian $H_{\text{eff}}$. We quantify the entanglement of a mixed state with its entanglement negativity $\mathcal{N}(\rho)=(\norm*{\rho^{\Gamma_A}}_1-1)/2$, where $\rho^{\Gamma_A}$ is the partial transpose of $\rho$ with respect to subsystem $A$~\cite{peres1996separability,eisert1999comparison,horodecki1996separability}. Using this metric, in the inset of \cref{fig:fin-temp}, we show that our similarity transformations indeed consistently reduce the entanglement of the Gibbs state. Furthermore, \cref{fig:off-diag} shows that, across a wide temperature range, our transformation's ability to suppress off-diagonal elements in the exact Gibbs state $\rho$ grows with $k$. This provides further evidence that Clifford + $k$Rz transformations provide an effective method to reduce the quantumness of a Hamiltonian, and moreover, that this reduction is tunable by varying $k$.

\textit{Conclusions.} We have developed a general approach for zero and finite temperature quantum calculations from a quantum information theoretic perspective, facilitated by the Clifford + $k$Rz transformation. One of the primary benefits of our approach is that it is fully classically simulable for small $k$, providing an convenient method to conserve costly quantum resources. Furthermore, by tuning $k$, we provide a means of customizing the similarity transforms produced by our technique: larger $k$ enables greatly increased expressive power and accuracy, at the cost of increased classical computation. This tunability forms the basis of what we term the quantum magic ladder. We apply these methods in the context of ground state and free energy calculations for quantum chemistry, demonstrating superior performance to HF techniques and consistent improvement as we ascend the quantum magic ladder, for both digital and analog quantum computers. We examine how the Clifford + $k$Rz transformation reduces quantumness (as measured by entanglement and magic) before any algorithms are run on quantum hardware, and demonstrated how this reduction directly translates to reduced requirements for quantum resources. While our approach shows promise for small-scale systems like LiH and H$_2$O, it is necessary to investigate its scalability to larger systems. The classical simulation cost of the Clifford+$k$Rz circuits grows exponentially with $k$, which could limit the applicability of our method to larger systems. However, it is encouraging to note that we are already able to find significant improvements for $k \leq 2$, which suggests that even a small number of strategically placed non-Clifford gates can provide significant benefits. As quantum hardware continues to improve, we expect that the insights and techniques developed in this work, such as the use of classically-optimized similarity transformations to reduce the ``quantumness" of the problem, will be valuable for scaling up to larger system sizes.

Having established this framework, we also identify a number of remaining open questions and future directions. Algorithmically, in addition to the greedy search algorithms used in this work, powerful methods such as simulated annealing and differentiable quantum architecture search~\cite{Zhang_2022} could be considered. Our similarity transformation based on doped Clifford circuits could also potentially augment well-known numerical approaches such as density matrix renormalization group (DMRG)~\cite{Schollw_ck_2005}. Since our method identifies effective Hamiltonians whose ground states have significantly lower entanglement compared to the original problem, it could potentially aid in the construction of more efficient tensor network representations. Moreover, recent work~\cite{masotllima2024stabilizer} has shown that the stabilizer formalism can be integrated with tensor networks. By transforming the problem into a basis where the ground state is closer to a stabilizer state, our approach could enable the use of these compact representations. Investigating the interplay between our similarity transformations and tensor network methods could lead to powerful new tools for simulating quantum many-body systems. Finally, to scale finite temperature calculations to larger systems, high-dimensional classical probability distributions of the Gibbs state can be simulated using generative models in the spirit of $\beta$-VQE~\cite{Liu_2021}. Our approach can also be applied to the finite temperature calculation based on the thermal double field framework \cite{takahashi1996thermo,wu2019}. This offers an opportunity to tailor our transformations to specific applications, opening new pathways for near-term quantum computers to address a wide array of quantum chemistry problems.

\begin{acknowledgments}
The authors would like to thank Chong Sun, Yi Tan, Linqing Peng, Yuan Liu, Shi-Xin Zhang, Joonho Lee, James Whitfield, and Ryan Babbush for the helpful discussions. DL acknowledges support from the NSF AI Institute for Artificial Intelligence and Fundamental Interactions (IAIFI). This material is based upon work supported by the U.S. Department of Energy, Office of Science, National Quantum Information Science Research Centers, Co-design Center for Quantum Advantage (C2QA) under contract number DE-SC0012704. HYH is grateful for the support from Harvard Quantum Initiative Fellowship. SFY would like to thank the NSF through a Cornell HDR and the CUA PFC grant.
\end{acknowledgments}

\bibliographystyle{abbrvnat}
\bibliography{refs}

\onecolumn
\newpage
\appendix
\section{\label{app:general-simulation}Classical simulation of Clifford-dominated quantum circuits}
In this section, we review the basic idea of the generalized stabilizer representation and how to simulate non-Clifford gates and general quantum states with stabilizer frames. First, let us define \emph{Pauli group} and \emph{stabilizer group}:
\begin{definition}[Pauli group]
    The \emph{Pauli group} on n-qubits is the set of all tensor products of Pauli matrices, i.e.
    \begin{equation}
        G_n = \{i^k \sigma^{i_1}\otimes \sigma^{i_2}\otimes \cdots \otimes \sigma^{i_{n}}|k,i_{h}\in\{0,1,2,3\}\}
    \end{equation}
\end{definition}

\begin{definition}[Stabilizer group]
    \emph{Stabilizer group} $\mathcal{S}$ is a subgroup of Pauli group satisfies the following properties:\\
        1. $\mathcal{S}$ is an abelian group. For $\forall s_i,s_j\in \mathcal{S}$, $[s_i,s_j]=0$. \\
        2. It doesn't contain negative identity, i.e. $-I \notin \mathcal{S}$.
\end{definition}
The stabilizer group can always be represented by its generators. For example, a stabilizer group containing $2^r$ elements can be represented by $r$ generators of the group: $\mathcal{S}=\langle s_1,s_2,\dots,s_r\rangle$. One should notice the generators are not unique. For example, $\mathcal{S}=\langle s_1,s_2,\dots,s_r\rangle=\langle s_1s_2,s_2,\dots,s_r\rangle$. With the stabilizer group, one can define stabilizer states:
\begin{theorem}[Pure stabilizer state]
    A full stabilizer (n generators for a n-qubit system) stabilizes a unique pure state $\ket{\psi_S}$
\end{theorem}
One can check the density matrix can be represented as 
\begin{equation}
    \rho_S=\dfrac{1}{2^n} \prod_{i=1}^{n}\left(I+s_i\right)
\end{equation}

Another concept that is useful for describing general quantum state is the \emph{destabilizers}:
\begin{definition}[Destabilizer group]
    A \emph{destabilizer group} $\mathcal{D}$ is a subgroup of Pauli group associated with the stabilizer group $\mathcal{S}=\langle s_1,s_2,\dots,s_r\rangle$. It has the following properties:\\
        1. $\mathcal{D}$ has the same number of generators as $\mathcal{S}$: $\mathcal{D}=\langle d_1,d_2,\dots,d_r\rangle$. \\
        2. For each generator $d_i$, it anti-commute with $s_i$, but commute with all the other generators $s_j,j\neq i$.
\end{definition}

\begin{definition}[Full tableau]
A full stabilizer tableau $\mathcal{T}=(\mathcal{S},\mathcal{D})$ is a pair of full stabilizer and destabilizer.
\end{definition}
\begin{definition}[Stabilizer basis]
    The stabilizer basis with full tableau $\mathcal{T}$ is $\mathcal{B}(\scT)=\{d\ket{\psi_S}|d\in\scD\}$
\end{definition}

The stabilizer basis is an orthonormal and complete basis for $n$-qubit Hilbert space. Since the destabilizers are note unique, if $\scD$ and $\scD'$ are both destabilizers for $\scS$, then $\scB(S,D)$ and $\scB(S,D')$ are the same set of basis modulo phase difference. Since $\scB(\scT)$ is a complete basis for Hilbert space, then any quantum state can be represented by the stabilizer basis. Therefore, we introduce the generalized stabilizer representation.

\begin{definition}[Generalized stabilizer representation]
    Any pure quantum state can be represented with stabilizer basis. More particularly, the density matrix $\rho$ of any pure quantum state can be written as 
    \begin{equation}
        \rho = \sum_{i=1}^{2^n}\sum_{j=1}^{2^n}\chi_{ij}d_i\ket{\psi_{S}}\bra{\psi_{S}}d_j = \sum_{ij}\chi_{ij}d_i \rho_{S}d_j.
    \end{equation}
    Therefore, we call the two-tuple $(\chi,\scB(\scS,\scD))$ the generalized stabilizer representation.
\end{definition}

For a general state, in the worst case, we need $\scO(4^n)$ classical memory to store and simulate its dynamics. However, this formulation is particularly useful when this decomposition is sparse. For simplicity, we can use the zero norm of $\chi$ matrix to denote its complexity, i.e. $\Lambda(\chi)=||\chi||_0$. And the simulation complexity scales as $\mathcal{O}(\Lambda(\chi))$. Now we define the \emph{Pauli channel} and see how to simulate general quantum states with the generalized stabilizer representation.
\begin{definition}[Pauli channel]
    The Pauli channel of a n-qubit system can be written as 
    \begin{equation}
        \scE(\rho)=\sum_{ij}\phi_{ij}P_i \rho P_j
    \end{equation}
    where $P_i,P_j\in G_n$, and $\phi_{ij}$ are complex numbers. We use $\Lambda(\scE)=||\phi||_0$ to denote the complexity of the Pauli channel.
\end{definition}
\subsection{General state evolution under Clifford gates}
The evolution of the general state under a Clifford circuit $C$ evolution can be written as
\begin{equation}
    C \rho C^{\dagger} = C\left(\sum_{ij}\chi_{ij}d_i \rho_S d_j\right)C^{\dagger} = \sum_{ij}\chi_{ij}C d_j C^{\dagger}\left(C \rho_S C^{\dagger}\right)C d_j C^{\dagger}
\end{equation}
Therefore, we only need to update the full tableau $\scT=(\scS,\scD)$, which can be done with $\scO(n^2)$ time. Under Clifford gates, the general stabilizer frame will update as $(\chi,\scS,\scD)\rightarrow(\chi,\scS',\scD')$.

\subsection{General state evolution under Pauli channels}

First, we use the fact that any Pauli string can be decomposed with stabilizer and destabilizers, $P_m=\alpha_{m}d_{b_{m}}s_{c_{m}}$. An efficient implementation of this decomposition is in \cref{algo:pauli-decom}.  Suppose we want to simulate a general Pauli channel, $\scE(\rho)=\sum_{m,n}\phi_{m,n}P_{m}\rho P_{n}^{\dagger}$. For each term in the Pauli channel:
\begin{equation}
    \begin{split}
        \phi_{m,n}P_m \rho P_n^{\dagger} &= \sum_{ij}\phi_{m,n}P_m \chi_{ij}d_i \rho_{S}d_j P_n^{\dagger}= \sum_{ij}\phi_{m,n}\chi_{ij} P_m d_i \rho_{S}d_j P_n^{\dagger}\\
        & = \sum_{ij}\phi_{m,n}\chi_{ij} \alpha_{m}\alpha^{*}_{n}d_{b_{m}} s_{c_{m}} d_i\rho_{S}d_j s_{c_{n}}d_{b_{n}}\\
        & = \sum_{ij}\phi_{m,n}\chi_{ij} \alpha_{m}\alpha^{*}_{n}(-1)^{c_m\cdot i+c_n\cdot j}d_{b_{m}}d_i s_{c_m}\rho_{S}s_{c_n}d_j d_{b_n}\\
        & = \sum_{ij}\left(\phi_{m,n}\chi_{ij} \alpha_{m}\alpha^{*}_{n}(-1)^{c_m\cdot i+c_n\cdot j}\right) d_{b_m + i}\rho_{S}d_{j+b_n}\\
        & = \sum_{i'j'}\chi'_{i'j'}d_{i'}\rho_{S}d_{j'}.
    \end{split}
\end{equation}
Therefore, for each term in Pauli channel, we need to update $(\chi,\scS,\scD)\rightarrow(\chi',\scS,\scD)$. And we sum over all the terms in the Pauli channel, which contains $\Lambda(\phi)$ terms. Especially, T-gate can be written as the following channel:
\begin{equation}
    \begin{split}
        T \rho T^{\dagger}=\left(\cos \dfrac{\pi}{8}I +i \sin\dfrac{\pi}{8}Z\right)\rho\left(\cos \dfrac{\pi}{8}I -i \sin\dfrac{\pi}{8}Z\right)
    \end{split},
\end{equation}
where $\Lambda(T)=4$.

\begin{algorithm}[H]
\caption{Pauli decomposition algorithm.}\label{algo:pauli-decom}
This algorithm efficiently decomposes a Pauli string $P$ into a product of stabilizer and destabilizer elements $\mathbf{d}_b,\mathbf{s}_{c}$, and a phase factor $\alpha$. The decomposition is used in the classical simulation of the Clifford+$k$Rz circuits.
\begin{algorithmic}[1]
  \Procedure{Decompose}{$P$}$\rightarrow \alpha, \mathbf{d}_b,\mathbf{s}_{c}$
    \State Initialize zero vector $\mathbf{b}=(0,\dots,0)$, and $\mathbf{c}=(0,\dots,0)$
    \State phase = 0
    \State $P_{t}=I$ \Comment{Binary vector encoding, no phase information.}
    \For{$i=0, \ldots, N-1$} \Comment{Loop over stabilizer generators}
        \If{$\{P,\mathbf{s}_i\}=0$}
        \State $b_i = 1$
        \State $\text{phase} \gets \text{phase} - \text{ipow}(P_t,\mathbf{d}_i)$
        \State $P_t=P_t*\mathbf{d}_i$
        \EndIf
    \EndFor
    \For{$i=0, \ldots, N-1$} \Comment{Loop over destabilizer generators}
    \If{$\{P,\mathbf{d}_i\}=0$}
    \State $c_i = 1$
    \State $\text{phase} \gets \text{phase} - \text{ipow}(P_t,\mathbf{s}_i)$
        \State $P_t=P_t*\mathbf{s}_i$
    \EndIf
    \EndFor
    \State \textbf{return} phase, $\mathbf{b}$, and $\mathbf{c}$.
  \EndProcedure
\end{algorithmic}
\end{algorithm}

\section{\label{app:optimization} Optimization of Clifford+\texorpdfstring{$k$}{k}Rz circuits}
\begin{figure}[ht]
    \centering
    \scalebox{1.0}{
\Qcircuit @C=1.0em @R=0.2em @!R { \\
	 	\nghost{{q}_{0} :  } & \lstick{{q}_{0} :  } & \multigate{1}{e^{i \pi P_0 /4}} & \push{\begin{tabular}{ |c| } $e^{i\pi P_4/4}$\\ \hline \end{tabular}} \qw \barrier[0em]{7} & \qw & \qw \barrier[0em]{7} & \qw & \multigate{1}{e^{i \pi P_8 /4}} & \push{\begin{tabular}{ |c| } $e^{i \pi P_{12}/4}$\\ \hline \end{tabular}} \qw & \qw & \qw\\
	 	\nghost{{q}_{1} :  } & \lstick{{q}_{1} :  } & \ghost{e^{i \pi P_0 /4}} & \multigate{1}{e^{i \pi P_5 /4}} & \qw & \qw & \qw & \ghost{e^{i \pi P_8 /4}} & \multigate{1}{e^{i \pi P_{13} /4}} & \qw & \qw\\
	 	\nghost{{q}_{2} :  } & \lstick{{q}_{2} :  } & \multigate{1}{e^{i \pi P_1 /4}} & \ghost{e^{i \pi P_5 /4}} & \qw & \gate{R_z(\theta_0)} & \qw & \multigate{1}{e^{i \pi P_9 /4}} & \ghost{e^{i \pi P_{13} /4}} & \qw & \qw\\
	 	\nghost{{q}_{3} :  } & \lstick{{q}_{3} :  } & \ghost{e^{i \pi P_1 /4}} & \multigate{1}{e^{i \pi P_6 /4}} & \qw & \qw & \qw & \ghost{e^{i \pi P_9 /4}} & \multigate{1}{e^{i \pi P_{14} /4}} & \qw & \qw\\
	 	\nghost{{q}_{4} :  } & \lstick{{q}_{4} :  } & \multigate{1}{e^{i \pi P_2 /4}} & \ghost{e^{i \pi P_6 /4}} & \qw & \qw & \qw & \multigate{1}{e^{i \pi P_{10} /4}} & \ghost{e^{i \pi P_{14} /4}} & \qw & \qw\\
	 	\nghost{{q}_{5} :  } & \lstick{{q}_{5} :  } & \ghost{e^{i \pi P_2 /4}} & \multigate{1}{e^{i \pi P_7 /4}} & \qw & \gate{R_z(\theta_1)} & \qw & \ghost{e^{i \pi P_{10} /4}} & \multigate{1}{e^{i \pi P_{15} /4}} & \qw & \qw\\
	 	\nghost{{q}_{6} :  } & \lstick{{q}_{6} :  } & \multigate{1}{e^{i \pi P_3 /4}} & \ghost{e^{i \pi P_7 /4}} & \qw & \qw & \qw & \multigate{1}{e^{i \pi P_{11} /4}} & \ghost{e^{i \pi P_{15} /4}} & \qw & \qw\\
	 	\nghost{{q}_{7} :  } & \lstick{{q}_{7} :  } & \ghost{e^{i \pi P_3 /4}} & \push{\begin{tabular}{ |c| }\hline $e^{i\pi P_4/4}$\\ \end{tabular}} \qw & \qw & \qw & \qw & \ghost{e^{i \pi P_{11} /4}} & \push{\begin{tabular}{ |c| } \hline $e^{i\pi P_{12}/4}$\\  \end{tabular}} \qw & \qw & \qw\\
\\ }}
    \caption{The brickwork circuit ansatz for $L=2$ and $k=2$.}
    \label{fig:ansatz}
\end{figure}
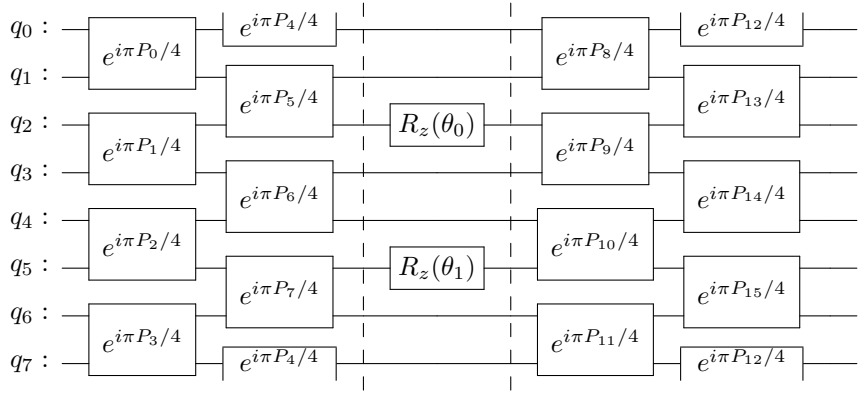

We use a brickwork circuit ansatz (see \cref{fig:ansatz}) with $L$ layers. One layer is composed of $n$ Clifford gates, each of which acts on neighboring qubits. Although a common gate set for the Clifford group is taken to be the Hadamard, phase, and CNOT gates, we instead use the gate set composed of $e^{i \pi P/4}$, where $P \in \{I,X,Y,Z\}^{\otimes 2}$. We dope the circuit with $k$ Rz gates at a fixed layer, but a variable site index (which is optimized over). To summarize, our free parameters belong to the domain $\mathbb{Z}_{16}^{n \times L} \times \mathbb{Z}_n^{k} \times \mathbb{R}^k$. The first term corresponds to 16 possible choices for the Clifford gates (of which there are $n \times L$). The second corresponds to the $n$ possible choices for the site index of the $k$ Rz gates, and the last is simply the argument for the $k$ Rz gates.

\begin{figure}[ht]
    \centering
    \includegraphics[width=0.65\textwidth]{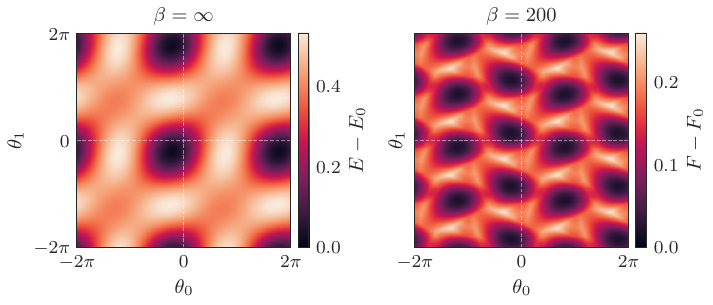}
    \caption{The landscape of $f(\Theta) \equiv f(X_{best},\Theta)$ for LiH at a bond length $3.4$\AA, where $X_{best}$ was the best performing solution for $k=2$. On the left, we show the landscape for the ground state calculation, and on the right, we show the landscape at a finite temperature $\beta=200$ ($T \approx 1500\text{K}$). Note that there are very few suboptimal local minima -- applying gradient descent to a randomly initialized $\Theta$ will find the global minimum with high probability.}
    \label{fig:theta-landscape}
\end{figure}

To optimize our circuit, we treat the discrete and continuous parameters separately. We label $X \in \mathbb{Z}_{16}^{n \times L} \times \mathbb{Z}_n^k$ the discrete parameters and $\Theta \in \mathbb{R}^k$ the continuous ones. For any cost function $f(X, \Theta)$, we can define a `marginalized' cost function by minimizing over $\Theta$:
\begin{equation}
    f(X) \equiv \min_{\Theta} f(X, \Theta).
\end{equation}
This is a useful construction when it is easy to do this minimization; it turns out that for small $k$, this process is very easy. Since the cost is periodic in $\Theta$, we only need to search the small volume $\Theta \in [0,2\pi]^k$. This search is further simplified by the fact that $f(X,\Theta)$ is continuous in $\Theta$, and does not have many local minima (see \cref{fig:theta-landscape}). By applying a simple gradient-based optimizer (e.g., gradient descent, conjugate gradient, etc.) to just a few ($\sim 10$) randomly initialized choices of $\Theta$, we can often find the global minimum within very few iterations. This is to say, the marginalized cost function $f(X)$ is easy to evaluate. With this, we optimize over the remaining parameters $X$ with a simple greedy search. Each iteration, we pick one of the parameters in $X$ to optimize over. For instance, if we choose to optimize $P_1$ (refer to \cref{fig:ansatz}), we would evaluate the cost with all other parameters held constant, varying $P_1$ over all 16 possibilities $\{I,X,Y,Z\}^{\otimes 2}$. Then, we simply replace $P_1$ with the choice that yielded the lowest cost $f(X)$. Finally, we do this discrete optimization in parallel for a large number ($n_{init} \sim 1000$) of randomly initialized parameters $X$, and simply take the best performing solution after a fixed number ($n_{iter} \sim 100$) iterations. Although we considered more advanced schemes for this discrete optimization, such as a simulated annealing process, we found that these schemes did not yield better results. 

\begin{algorithm}[H]
\caption{Optimizing Clifford + $k$Rz Circuits}\label{alg:cap}
\begin{algorithmic}
\Procedure{Optimize}{$f,L,k,n_{iter}$}
\State $X \gets$ random initialization in $\mathbb{Z}_{16}^{n \times L} \times \mathbb{Z}_n^k$
\For{$i=1,\ldots,n_{iter}$}
\State $j \gets$ random index in $1, \ldots, n \cdot L + k$
\If{$j \leq n \cdot L$} 
    \State $params \gets \mathbb{Z}_{16}$
\Else
    \State $params \gets \mathbb{Z}_n$
\EndIf 
\For{$p \in params$}
\State $X[j] \gets p$
\State $\text{cost}[p] \gets f(X)$
\EndFor
\State $X[j] \gets \text{argmin}_p \text{cost}[p]$
\EndFor
\EndProcedure
\end{algorithmic}
\end{algorithm}

\section{Symmetry of the Recovered Solutions\label{app:symmetry}}
\begin{figure}
 \centering
 \includegraphics[width=\textwidth]{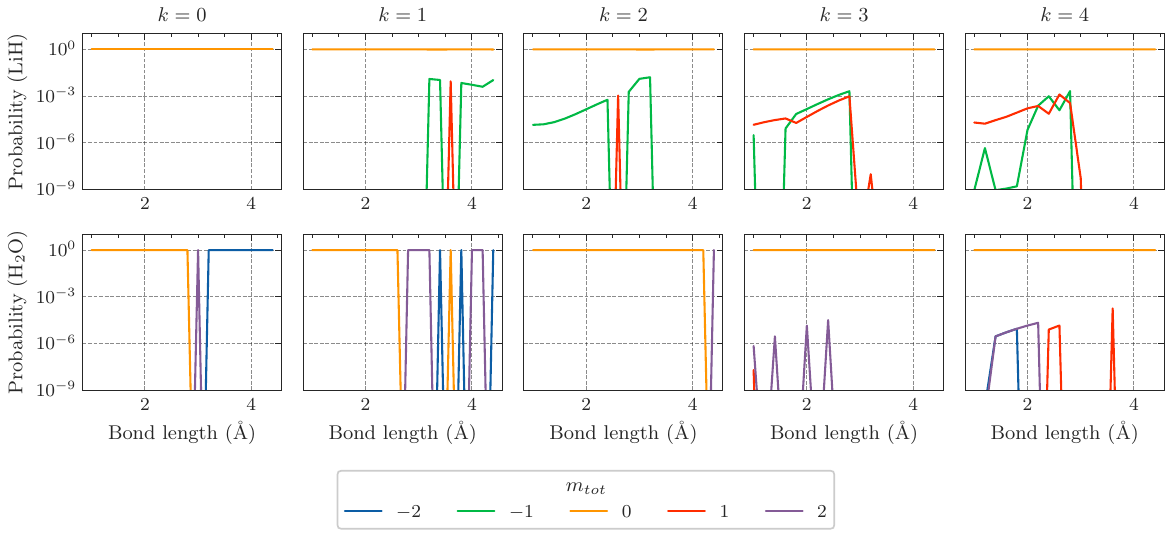}
 \caption{We analyze optimized Clifford + $k$Rz solution in the eigenbasis for the qubitized spin operator $S$ for both LiH (top) and H$_2$O (bottom), demonstrating that the recovered solution, for most cases, approximately conserves of $S$.}
 \label{fig:spin}
\end{figure}

A noteworthy drawback of the Clifford + $k$Rz ansatz (\cref{fig:ansatz}) is that it does not necessarily conserve any symmetries of the Hamiltonian. Despite this, we observe that the optimization process \textit{prefers} to recover solutions that are both eigenstates of the number and spin operators, and furthermore, typically states that conserve symmetries in the Hamiltonian (e.g., total spin zero). Although observe that the latter does not hold for every solution (see \cref{fig:spin}), it holds in the vast majority of cases. Remarkably, we observe a greater tendency to preserve symmetries as more Rz gates are injected. However, we did not explicitly investigate the relationship between these preserved symmetries and the magic of the initial state. Understanding this connection could provide valuable insights into the role of symmetries and magic in our approach. Future work could explore this relationship in more detail, potentially leading to better ansatz designs and improved performance.

\begin{figure}
 \centering
 \includegraphics[width=\textwidth]{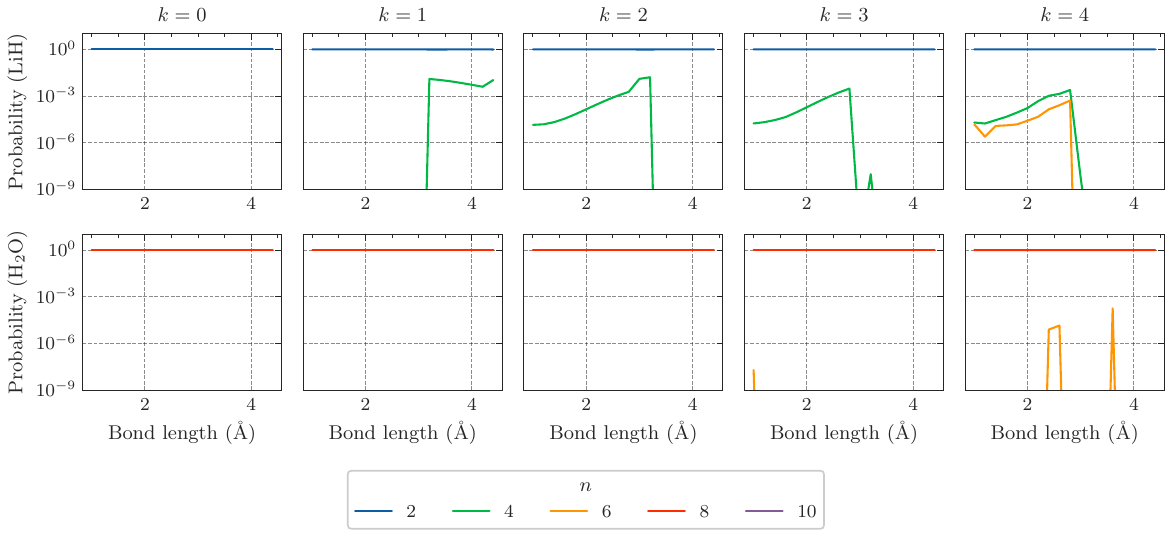}
 \caption{We analyze optimized Clifford + $k$Rz solution in the eigenbasis for the qubitized number operator $N$ for both LiH (top) and H$_2$O (bottom), demonstrating that the recovered solution approximately conserves of $N$.}
 \label{fig:num}
\end{figure}

\section{\label{app:Rydberg}Pulse optimization of programmable quantum simulators}
The Rydberg Hamiltonian can be modeled as 
\begin{equation}
H/\hbar=\sum_{j}\dfrac{\Omega_j(t)}{2}\left(e^{\ii \phi_j(t)}\ket{g}\bra{r}+e^{-\ii \phi_j(t)}\ket{r}\bra{g}\right)-\sum_{j}\Delta_j(t)\hat{n}_j+V_{ij}\hat{n}_{i}\hat{n}_j,    
\end{equation}
where $\Omega_j(t)$, $\phi_j(t)$, $\Delta_j(t)$ are Rabi frequency, laser phase, and local detuning of the driving field for each atom, and $\ket{g}$ is the ground state, $\ket{r}$ is the Rydberg state, $\hat{n}_j=\ket{r}_j\bra{r}_j$ is the number operator. We use $V_{ij}=C_6/|x_i-x_j|^6$ with $C_6=2\pi*8.627*10^5 ~\text{MHz}~\mu m^6$ to describe the Rydberg interactions. In our optimization, we translate the Hamiltonian into Pauli basis, which gives
\begin{equation}
    H/\hbar=\sum_j \left(\dfrac{\Omega_j}{2}\cos \phi_j X_j-\dfrac{\Omega_j}{2}\sin \phi_j Y_j\right)+\sum_{j}\left(\dfrac{\Delta_j}{2}-V_{ij}\right)Z_j+\sum_{i,j}\dfrac{V_{ij}}{4}Z_i Z_j+H_{\text{const}},
\end{equation}
where $H_{\text{const}}=\sum_{i,j}\frac{V_{ij}}{4}I-\sum_j \frac{\Delta_j}{2}I$. Then we optimize the pulse in the Pauli basis, such as $\frac{\Omega_j}{2}\cos \phi_j$. We parameterize each pulse as a piece-wise constant function with ten pieces, and initialize them randomly. The atom positions are fixed to be separated with 9.37$\mu m$. Since the interaction strength of the next nearest neighbors are 64 times smaller than the interaction strength of the nearest neighbors, we only consider the leading nearest neighbor interactions for our model.

\section{\label{app:finite-temp}Finite temperature calculation}
The variational principle extends beyond ground state calculations; indeed, it can be shown that the free energy satisfies a variational principle. That is, defining
\begin{equation}
    F_\beta(\rho) \equiv \Tr(H \rho) - S(\rho)/\beta,
\end{equation}
where $S(\rho)$ is the von Neumann entropy, it can be shown that for all $\beta$,
\begin{equation}
    F_\beta(\rho) \geq F_\beta\qty(\frac{e^{-\beta H}}{\Tr(e^{-\beta H})}) \equiv F_0
\end{equation}
for all $\rho$. We take advantage of this variational principle to estimate the free energy at finite temperatures. The procedure is simple: we fix an orthogonal basis of stabilizer states $\qty{\ket{\psi_k} \mid k=1,\ldots,2^n}$ (typically derived from SBRG), so that the initial state is
\begin{equation}
    \rho_0 \equiv \sum_k p_k \ketbra{\psi_k},
\end{equation}
where the $p_k \in [0,1]$ form a probability distribution (i.e., $\sum_k p_k = 1$). We construct a Clifford+kRz unitary $U$ from Clifford + $k$T gates, and evolve each of the  under this unitary. The final state is therefore
\begin{equation}
    \rho = \sum_k p_k U\ketbra{\psi_k} U^\dagger.
\end{equation}
Defining $E_k \equiv \Tr((H-\mu N) U \ketbra{\psi_k} U^\dagger)$, we observe that
\begin{equation}
    F_\beta(\rho) = \sum_k p_k \qty(E_k-\frac{\log p_k}{\beta}).
\end{equation}
The optimal $p_k$ can be solved for in closed form (i.e., they do not need to be optimized numerically), yielding a free energy estimate
\begin{equation}
    F_\beta(\rho) = -\frac{1}{\beta} \log(\sum_k \exp(-\beta E_k)).
\end{equation}
Since calculating $E_k$ as a function of the unitary $U$ is efficient, we can then apply our optimization techniques over the unitary $U$ to minimize $F_\beta(\rho)$.

\end{document}